\documentstyle[psfig]{mn}

\def\etal{{\it et al. }}
\title[NGC~3379 Globular Clusters]
{UBRI Photometry of Globular Clusters in the Leo Group
Galaxy NGC~3379}

\author[Whitlock, Forbes \& Beasley]{Shannon Whitlock$^{1}
$\thanks{swhitlock@swin.edu.au}, Duncan
A. Forbes$^{1}$\thanks{dforbes@astro.swin.edu.au},
Michael A. Beasley$^{1}$\thanks{mbeasley@astro.swin.edu.au}\\
  $^1$ Centre for Astrophysics \& Supercomputing, Swinburne University,
  Hawthorn, VIC 3122, Australia\\
}
\begin{document}
\maketitle

\begin{abstract}

We present wide area
UBRI photometry for globular clusters around the Leo group galaxy
NGC~3379. Globular cluster candidates are selected from their B-band
magnitudes and their (U--B)$_o$ vs (B--I)$_o$
colours. A colour-colour selection region was defined from
photometry of the Milky Way and M31 globular cluster systems.
We detect 133 globular cluster candidates which, supports previous
claims of a low specific frequency for NGC~3379.

The Milky Way and M31 reveal blue and red
subpopulations, with (U--B)$_o$ and
(B--I)$_o$ colours indicating mean metallicities
similar to those expected based on previous spectroscopic work.
The stellar population models of Maraston (2003) and Brocato
\etal (2000) are
consistent with both subpopulations being old, and with
metallicities of [Fe/H] $\sim$ --1.5 and --0.6 for the
blue  and red  subpopulations
respectively. The models of Worthey (1994) do not reproduce the
(U--B)$_o$ colours of the red (metal-rich) subpopulation
for any modelled age.

For NGC~3379 we detect a blue subpopulation with similar colours
and presumably age/metallicity,
to that of the Milky Way and M31 globular cluster systems. The
red subpopulation is less well defined, perhaps due to
increased photometric errors, but indicates a mean
metallicity of [Fe/H] $\sim$ --0.6.

\end{abstract}


\section{Introduction}\label{sec_intro}

In the last few years, it has become clear that globular cluster (GC)
systems have complex colour distributions, indicating two or more
subpopulations within a single elliptical galaxy (Ashman \& Zepf
1992; Secker \etal 1995; Whitmore \etal 1995; Geisler \etal 1996;
Forbes, Brodie \& Huchra 1997; Bridges \etal 1997; Kissler-Patig
\& Gebhardt 1998; Kundu \& Whitmore 2001; Larsen \etal 2001).
The subpopulations have different metallicities and possibly ages,
indicating multiple epochs or mechanisms of formation.  As GCs
are thought to
trace the star formation and chemical enrichment episodes of
their host galaxy (e.g. Forbes \& Forte 2001),
understanding how they formed will provide important constraints
on the process of galaxy formation and evolution. The main
scenarios for GC formation include:\\

\noindent
{\bf $\bullet$} The merger
of two gas--rich (spiral) galaxies may lead to the formation of an
elliptical galaxy and create an additional population of GCs in
the process (Ashman \& Zepf 1992). Since the GCs
produced in the merger formed from
enriched gas they should be of higher metallicity and
thus redder than the indigenous (metal-poor) GC population.
Thus we expect a metal-poor old population ($\sim$13 Gyrs)
plus a metal-rich young
population with an age similar to that of the merger itself.\\

\noindent
{\bf $\bullet$} A
multi--phase collapse (Forbes, Brodie \& Grillmair 1997)
can also produce two distinct GC populations. Here the blue GCs
formed in an early chaotic phase of galaxy formation from metal-poor gas
and the red GCs later from enriched gas
in the same
phase that produces the bulk of the galaxy starlight.
A multi--phase collapse also predicts an old metal-poor subpopulation and
one a slightly younger ($\Delta$ age $\sim$2--4 Gyr) metal-rich one.
\\

\noindent
{\bf $\bullet$} Cot\'e,
Marzke \& West (1998) describe the build-up of the GC systems of bright
ellipticals via the accretion of mostly metal-poor GCs from dwarf
galaxies. In this picture the metal-rich GCs are indigenous and
the metal-poor ones are acquired. Here both the GC subpopulations
will be about the same age (i.e. old) but with different metallicities.
\\

\noindent
{\bf $\bullet$} Formation of GC systems
in hierarchical merging is described
in Beasley \etal (2002). In this prescription blue GCs form in
pre-galactic clumps. These gaseous clumps merge generating a
second generation of red GCs, along with galactic star
formation. Late stage mergers of more stellar
clumps may include accreted GCs. Thus the hierarchical picture
contains elements of the other three. In a hierarchical Universe,
the metal-poor GCs will
be old with the metal-rich GCs having a mean age that
depends on galaxy mass and environment.\\

Perhaps the best direct test of these competing GC formation
models is to determine the mean age and metallicity of the GC
subpopulations for a large number of galaxies.
In principle, the best way to do this is from GC spectra.
Indeed this is an active area of research using the
world largest telescopes, and is returning exciting new results
(e.g. Kissler-Patig \etal 1998;
Forbes \etal 2001; Larsen \etal 2002; Beasley \etal 2003).
However it is also very time consuming. Photometry is
more efficient, but optical colours suffer from the well-known
age-metallicity degeneracy. This situation can be improved by
extending photometry to the near-infrared (e.g. Puzia \etal 2002)
or ultra-violet. Photometry in the U (3600\AA) band is very rare
for GC systems beyond the Local Group (the $\sim$70 GCs in
NGC~5128 with U-band photometry from Rejkuba 2001 is one exception).
This is largely due to the poor blue
response of most CCDs in use today and the low fraction of U-band
light emitted by old stellar populations such as GCs.

Here we present UBRI photometry of GCs
associated with the Leo Group galaxy NGC~3379,
obtained with the blue
sensitive CCDs of the Isaac Newton Telescope 2.5~m Wide Field Camera.
By extending the traditional
optical photometry to bluer U-band wavelengths, we can better
probe the metallicity distribution of the GC system in NGC~3379.
We also utilise some smaller field-of-view images taken with the
Gemini North 8m telescope to aid in the selection of GC
candidates.

NGC~3379 (M105) is a moderate luminosity E1 galaxy in the nearby
(D = 11.5 Mpc, m--M = 30.30) Leo Group. In a photographic study of
NGC~3379, Harris \& van den Bergh (1981) estimated a total GC
population of 290 $\pm$ 150. This translates into a low specific
frequency S$_N$ of 1.1 $\pm$ 0.6 (assuming M$_V$ = --21.06). Ajhar
\etal (1994) obtained VRI CCD images of NGC~3379 detecting some 60
GCs. However, they did not detect any obvious bimodality in the GC
colour distribution. The first clear detection of bimodality came
from the HST study by Larsen \etal (2001). They found peaks at
(V--I)$_o$ = 0.96 and 1.17 for the blue and red subpopulations
respectively.
Throughout this paper we adopt $H_o=75\,\mathrm{km\,s^{-1}\,Mpc^{-1}}$.

\section{Observations and data reduction}\label{sec_obs}

Broadband UBRI images covering the Leo galaxies NGC 3379, NGC
3384 and NGC 3389 were obtained using the 2.5~m Isaac Newton
Telescope (INT) on 2000 February 6th and 8th. The Wide Field
Camera (WFC) comprises 4 thinned EEV 4096x2048 CCDs, with
pixels of 0.33$^{''}$ and provides a field-of-view of approximately 30
$\times$ 30 arcmin.
Observing conditions were photometric over the two nights with
seeing of 1.5$^{''}$ in U and I bands and 2.2$^{''}$ in B and R
bands. The total exposure times were roughly
2000~s (U), 3000~s (B), 1500~s (R) and 1000~s (I).
In addition to the galaxy observations, several standard star fields
from Landolt (1992) were obtained over both nights which bracket
the galaxy observations.

Basic data reduction was performed using IRAF and specially
written software by A. Terlevich. This consisted of master bias
subtraction, non-linear correction, flat-fielding using combined
sky flats, alignment and co-addition of individual frames. The
galaxy frames and standard stars were reduced in an identical
manner. Raw magnitudes of between 10 and 30 stars were obtained
for each filter using the IRAF task QPHOT, after determining the
optimal aperture size. The zero point for each filter was
determined by a simple linear fit to the stellar raw magnitudes
versus stellar colour and a correction for airmass.  The airmass
extinction coefficients of K$_U$ = 0.46, K$_B$ = 0.22, K$_R$ =
0.08 and K$_I$ = 0.04 mag/airmass have been taken from the INT WFC
web page. The final airmass corrected, one second zero points are
Z$_U$ = 22.98 $\pm$ 0.01, Z$_B$ = 24.84 $\pm$ 0.02, Z$_R$ = 24.65
$\pm$ 0.02 and Z$_I$ = 23.96 $\pm$ 0.01. The BRI zero points
compare well with those determined by Mills et al. (2003) from the
same observing run and the zero points listed on the INT/WFC web
page. An independent estimate of the U band zero point (Z$_U$ =
22.98 $\pm$ 0.05) was made by F. Reda (2002, priv. comm.) by
comparing various aperture magnitudes for NGC 3379 to that listed
in Hypercat (http://www-obs.univ-lyon1.fr/hypercat/). Finally, we
adjusted these zero points for Galactic extinction using values
from the NASA Extragalactic Database
(http://ned.ipac.caltech.edu), i.e. A$_U$ = 0.132, A$_B$ = 0.105,
A$_R$ = 0.065, A$_I$ = 0.047.

\section{Galaxy Modelling}

In order to better reveal the inner GCs, we subtracted a model of
NGC~3379.
Galaxy subtraction was performed using the STSDAS ISOPHOTE package
(see Forbes \& Thomson 1992). A galaxy model, with a varying centre,
ellipticity and position angle, was fit
in all four filters. Isophotes were modelled out to
$\sim$350$^{''}$ (20 kpc).
A 3 sigma-clip criterion over 5 iterations was used to
remove deviant pixels (e.g. bright objects) from the fit. The
model-subtracted images were visually inspected and iterated if
necessary to provide a smooth transition from the background level to
the model-subtracted region. The resulting residual images made
it easier to identify the inner GCs.

\begin{figure}
\centerline{\psfig{figure=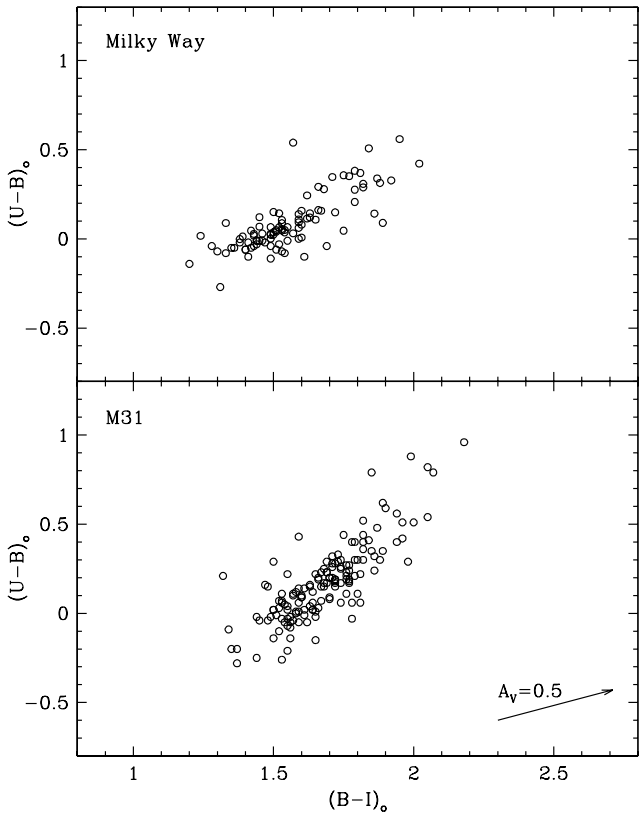,width=0.5\textwidth,angle=0}}
 \caption
{(U--B)$_o$ vs (B--I)$_o$ colour distribution for the Milky
Way (top) and M31 (lower) GCs. The distribution
for the two galaxies are similar, although they have slightly
different slopes.
The effect of 0.5$^m$
of extinction in the V band is also shown.
}
\label{double}
\end{figure}

\begin{figure}
\centerline{\psfig{figure=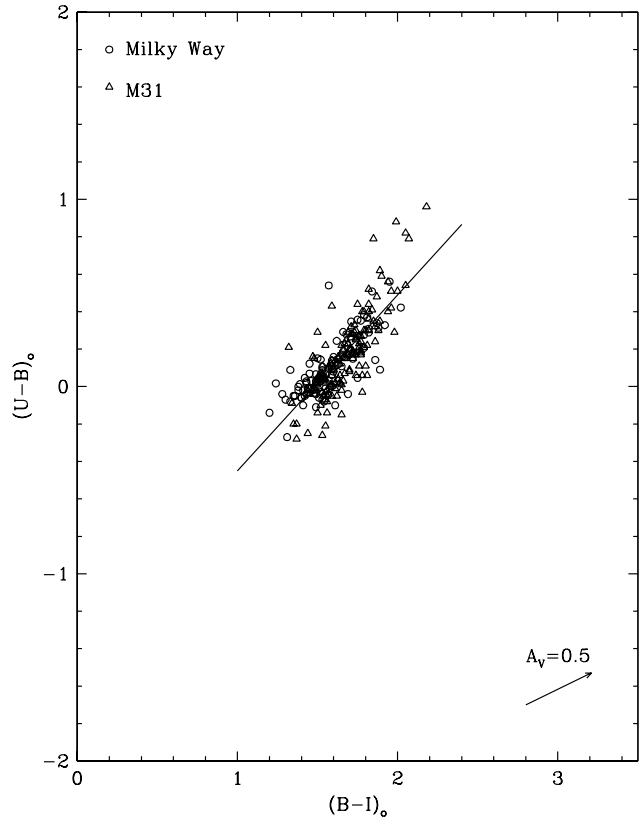,width=0.5\textwidth,angle=0}}
 \caption
{
(U--B)$_o$ vs (B--I)$_o$ colour distribution for the combined Milky
Way (circles) and M31 (triangles) GC samples.
A least squares fit to the combined sample is also shown. The
best fit line is of the form (U--B)$_o$ = 0.94 $\times$ (B--I)$_o$
- 1.39, with an rms spread about the best fit of
$\sim$ 0.12 mag.
}
\label{combine}
\end{figure}

\section{Initial Object Finding and Selection}

The four filter (i.e. UBRI) images of the Leo galaxy triplet
were aligned to within a
fraction of a pixel and trimmed using the IMALIGN task. We then used
DAOFIND to select GC candidates in each filter independently.
Selection criteria consisted of a S/N threshold of 6 and PSF FWHM
adjusted to the seeing conditions in each filter.  Additionally,
roundness criteria ranging between values of -1.0 and 1.0, and a
sharpness range of 0 to 1.0 provided good exclusion of extended
objects.  The resulting database of GC candidates consisted of
over 3,000 objects for each filter.
The PHOT task was used to
measure the magnitude and error of each object in all four bands
as found by DAOFIND.  A simple script was written to perform the
first stage of filtering and reduce the number of candidates
to only those detected in all four bands.  This spatial
coordinate matching reduced the dataset to 1,200 objects with
measured UBRI magnitudes.
The same script determined galactocentric coordinates for each
object based on the galaxy centre taken from the galaxy model
(with the centre position constant
within a fraction of a pixel).  The reduced candidate  list was then
visually inspected to eliminate the small number of
remaining galaxy-like extended
objects or CCD artifacts.

\section{Milky Way and M31 Globular
Cluster Colours}

To aid in the selection and interpretation of Leo group GCs, we
use observations of Milky Way and M31 GCs (essentially
the only two galaxies that have well-studied GC systems in
the U-band). Data for the Milky Way GC system comes from the
compilation of Harris (1996). From this list de-reddened U, B,
and I-band photometry exists for 95 GCs. Data for the M31 GC system
come from Barmby \etal (2000). De-reddened magnitudes for 148 GCs
were supplied by Barmby (2003, priv. comm.). In Fig.~\ref{double} we show
the (U--B)$_o$ vs (B--I)$_o$ colour distribution for the
Milky Way and M31 GC systems.
To our knowledge this is the first time the
(U--B)$_o$ vs (B--I)$_o$ colour distribution for the GCs of these
galaxies have been examined together.

\begin{figure*}
\centerline{\psfig{figure=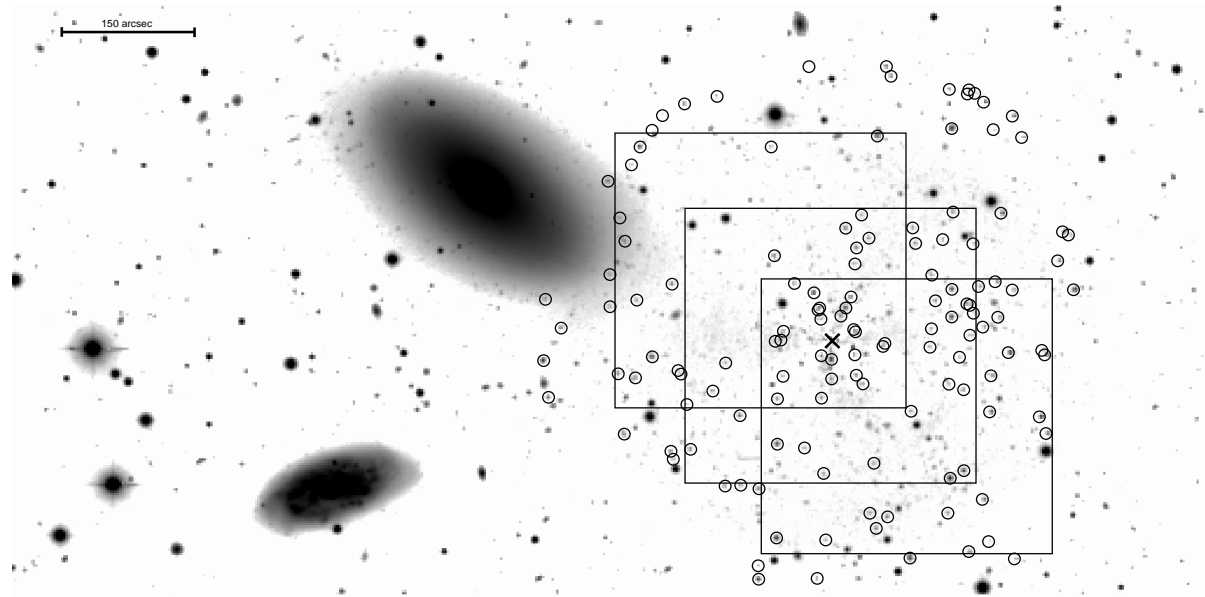,width=0.9\textwidth,angle=0}}
 \caption
{Grey scale image of the Leo triplet field from the INT/WFC.
NGC~3379 has been modelled and subtracted with its centre
indicated by a cross. Small circles represent globular
cluster candidates. The three large squares indicate the three
regions covered by the Gemini/GMOS imaging. North is up and East
is left. A 150 arcsec (8.4 kpc) size scale is shown.
}
\label{field}
\end{figure*}

\begin{figure}
\centerline{\psfig{figure=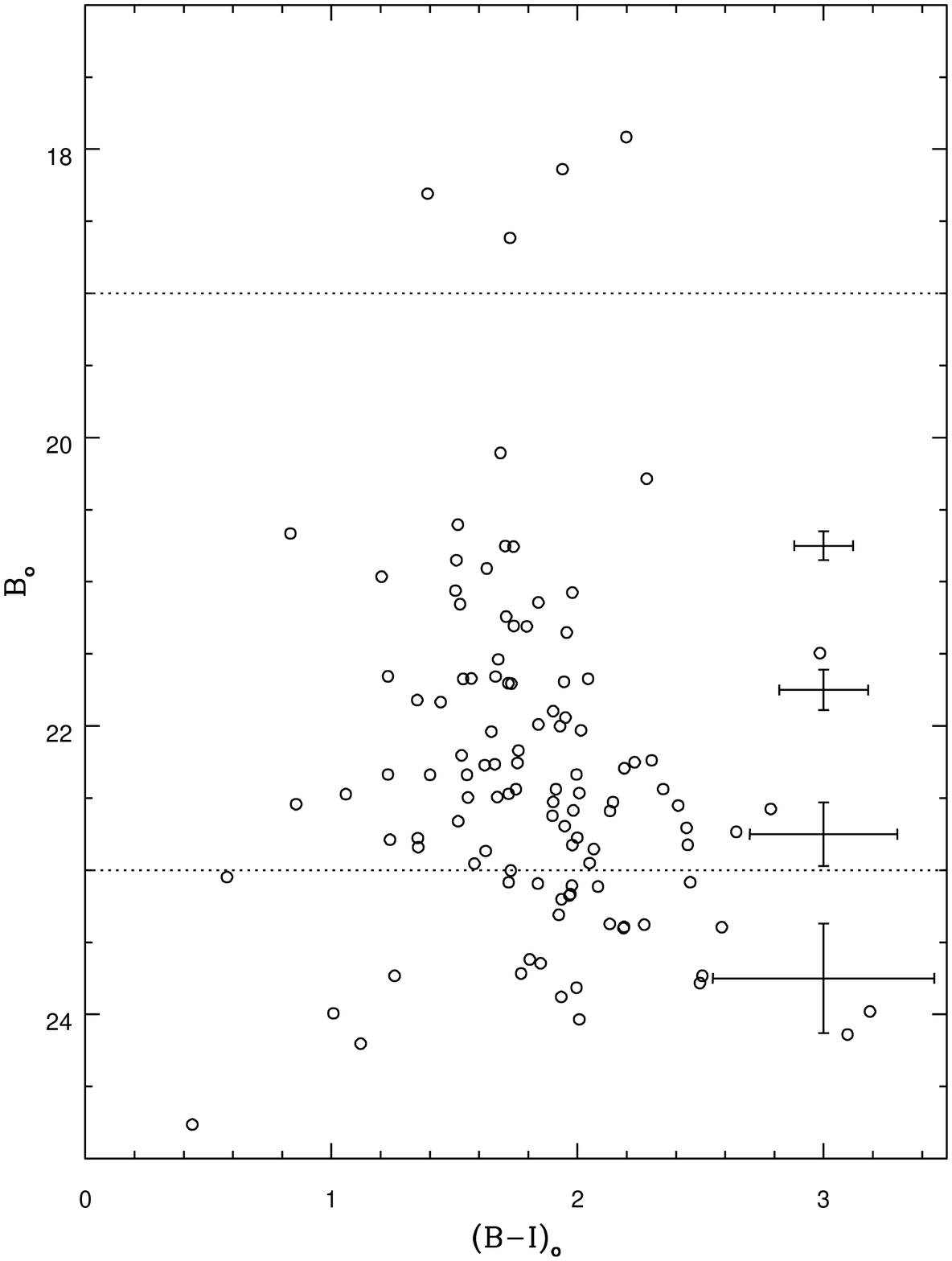,width=0.5\textwidth,angle=0}}
 \caption
{Colour-magnitude diagram for the candidate GCs from Gemini/GMOS
images, using magnitudes from the INT/WFC. The expected GC
turnover magnitude for NGC~3379 at a distance modulus of
m--M = 30.30 is B$_o$ = 23.5. The data show a
hint of bimodality at (B--I)$_o$ $\sim$ 1.6 and 1.9. The dashed lines
indicate the upper and lower magnitude cuts applied to this sample.
}
\label{cmd}
\end{figure}

The data define a relatively tight distribution in this plot
covering a range of $\sim$1.5 mags in each colour, with the M31 GCs
extending to slightly redder colours. In general, the
distributions for the two galaxies are qualitatively
similar. However we do note that the slope of the Milky Way
distribution (U--B)$_o$ = 0.725$\times$(B--I)$_o$--1.047 (rms = 0.088),
is somewhat flatter than that for M31
(U--B)$_o$ = 1.202$\times$(B--I)$_o$--1.850 (rms = 0.122).
This may represent intrinsic differences in the GC systems
of the two galaxies, but is more likely to be due to the
uncertain reddening corrections that have been applied.
Both galaxies reveal a dominant grouping of blue GCs,
with a less well-defined tail to redder
colours. We associate the former with the
metal-poor GC subpopulation and the latter with the
metal-rich GC subpopulation in these galaxies. The location of
these subpopulations in colour space will be discussed in more
detail below. Noting the general similarities of the two distributions,
we have combined the Milky Way and M31 GC samples to
define a mean GC colour-colour distribution.
This is shown in Fig.~\ref{combine}.
A least squares fit to the combined sample is also shown. The
best fit line is of the form (U--B)$_o$ = 0.94 $\times$ (B--I)$_o$
- 1.39, with an rms spread about the best fit of
$\sim$ 0.12 mag. We use this colour-colour relation and spread
to help us define the colours expected for GCs in NGC~3379.

\begin{figure}
\centerline{\psfig{figure=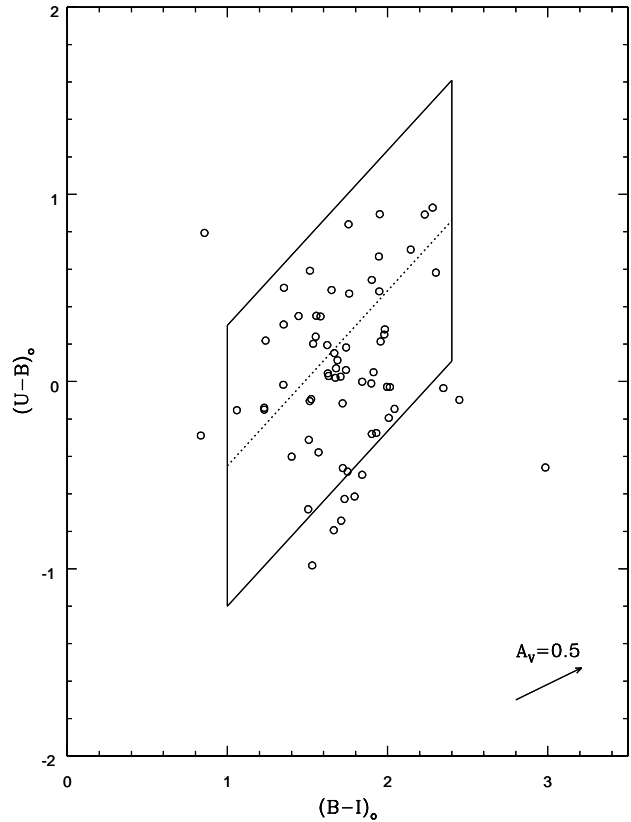,width=0.5\textwidth,angle=0}}
 \caption
{(U--B)$_o$ vs (B--I)$_o$ colour distribution for the candidate
GCs selected
from Gemini/GMOS images,
after applying spatial matching and B magnitude cuts.
Most of the data lie in a parameter space defined from the M31
and Milky Way distribution, indicating that the contamination
from background objects in
the Gemini data is indeed quite small. The data suggest a clear
blue subpopulation with colours (U--B)$_o$ $\sim$ 0.1 and
(B--I)$_o$ $\sim$ 1.65. A red
subpopulation is not clearly identified in this colour-colour diagram.
}
\label{geminiubi}
\end{figure}


\section{Globular Clusters in the Central Region of NGC~3379}

Images of the central regions of NGC~3379 have been taken with
the GMOS instrument on the Gemini North Telescope in 2003
February. Three fields, each covering 5 arcminutes near the galaxy
centre, were observed as shown in Fig.~\ref{field}. 
These images, in Sloan filters g$^{'}$, r$^{'}$ and i$^{'}$,
form the pre-acquisition imaging for an upcoming spectroscopic
run. As they were obtained under excellent seeing conditions
($\sim$0.7$^{''}$) the expected contamination rates for candidate
GCs based on the Gemini data will be significantly less than for the
INT data (obtained under $\sim$2$^{''}$ seeing). Candidate GCs
from the Gemini data have been selected on the basis of their
size (i.e. compactness) and
colour ($0.5 < g^{'}-i^{'} < 1.5$ and $0.3 < g^{'}-r^{'} < 1.0$) by
Faifer \& Forte (2003, priv. comm.).

By spatially matching the Gemini object list with the initial INT
object list (described above), we identified 125 matches.
In Fig.~\ref{cmd} we show a colour-magnitude
diagram for these Gemini-selected objects, with magnitudes taken
from our INT photometry.
Next we restricted the object list in B magnitude. An upper limit
of B = 19 was chosen to exclude bright stars and/or compact dwarf
galaxies. This limit is 4$\sigma$ brighter than the expected GC
turnover magnitude for NGC~3379, and corresponds to M$_B$ =
--11.3.
We also imposed a lower magnitude cutoff of B = 23,
to avoid any colour bias in the sample. These selection criteria
are also shown in Fig.~\ref{cmd}.

\begin{figure}
\centerline{\psfig{figure=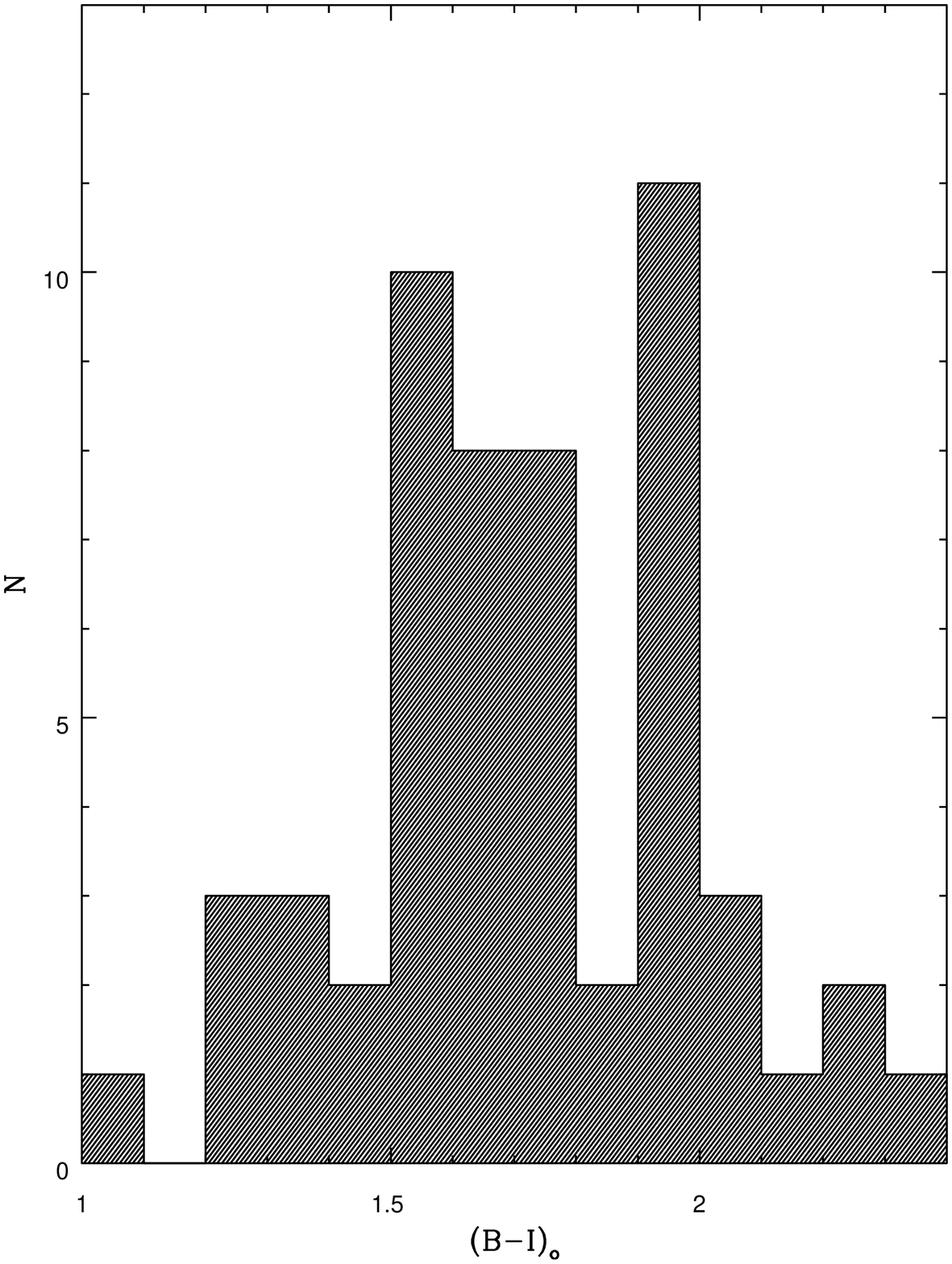,width=0.5\textwidth,angle=0}}
 \caption
{Histogram of (B--I)$_o$ colours for candidate GCs after
colour-colour selection from Gemini/GMOS imaging.
Two clear peaks are seen. A KMM statistical test
confirms peaks at (B--I)$_o$ = 1.65 and 1.90 with similar number
of GCs in each subpopulation.
}
\label{geminihist}
\end{figure}

The resulting subset in a (U--B)$_o$ vs (B--I)$_o$ colour-colour
diagram is shown in Fig.\ref{geminiubi}. This figure also shows
the region of expected colours for
GCs. The region ranges from 1.0 $<$ (B--I)$_o$ $<$ 2.4, and within
0.7 magnitudes in (U--B)$_o$ of the mean Milky Way plus M31 fit. The
(B--I)$_o$ range
was chosen to be similar to that of the Milky Way and M31, i.e.
covering the full metallicity range expected of GCs but
allowing for an additional 0.25 mag in A$_V$
reddening. The range in (U--B)$_o$
corresponds
to the scatter seen in the combined Milky Way and M31 datasets of
0.12 mag, added in quadrature with our typical (U--B)$_o$
photometric error.
Most of the data lie in the defined region,
indicating that the contamination in
the Gemini data is indeed low. The data show a group of blue GCs with
values (U--B)$_o$ $\sim$ 0.1 and (B--I)$_o$ $\sim$ 1.65. A second,
red subpopulation is difficult to clearly identify.


If we examine only the (B--I)$_o$
colours of the objects within the colour selected region,
we then obtain the histogram shown in
Fig.~\ref{geminihist}.
Visually, and via a KMM statistical test (Ashman, Bird \& Zepf
1994), the Gemini GC candidates are clearly bimodal. The
peaks are located at (B--I)$_o$ = 1.65 and 1.90.

\section{Large Area Study of the NGC~3379 Globular Cluster System}

\begin{figure}
\centerline{\psfig{figure=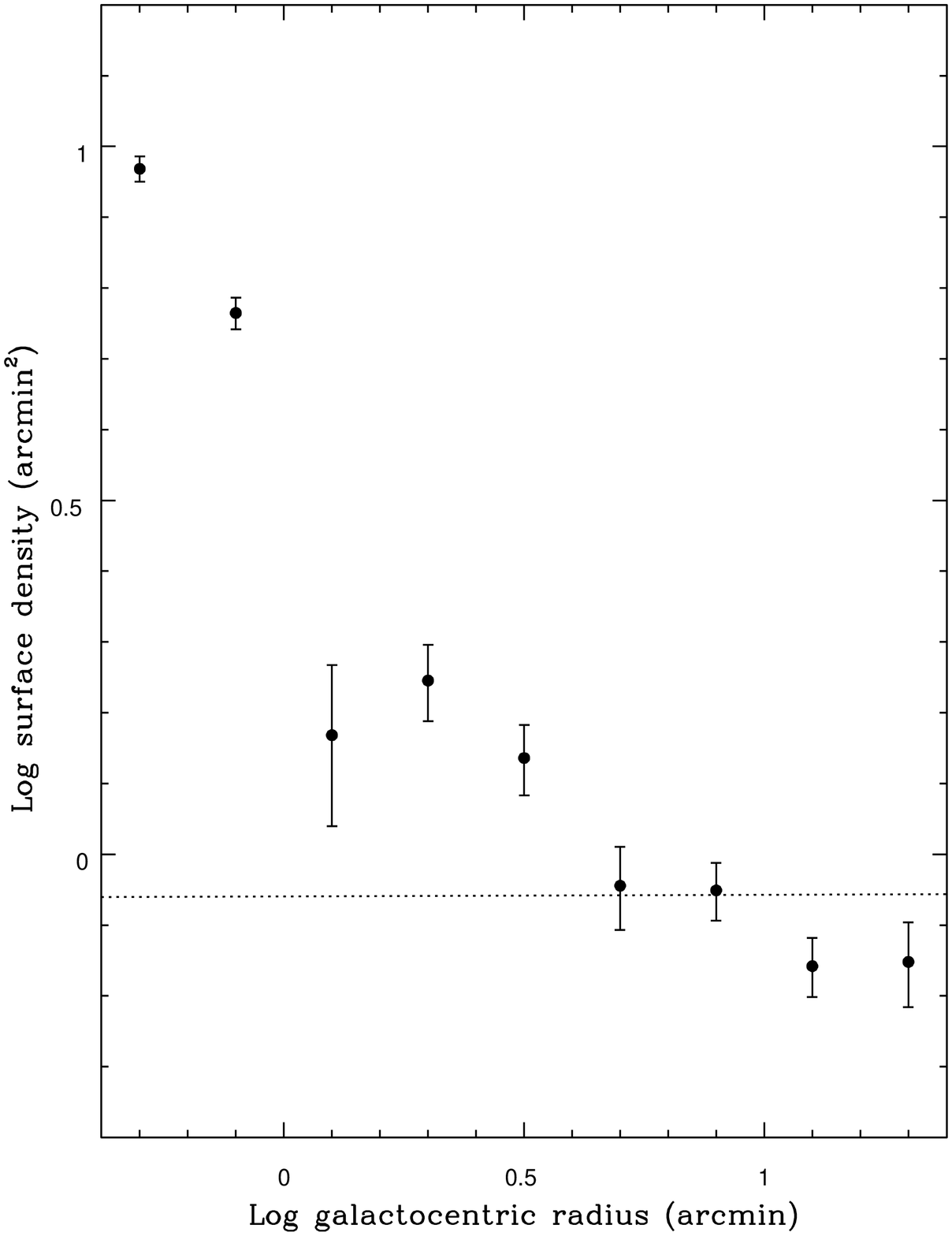,width=0.5\textwidth,angle=0}}
\caption {Globular cluster surface density against galactocentric 
radius for the INT/WFC sample. The dotted line indicates
a background region surface density at a radius of greater than 13$^{'}$. 
} \label{radial}
\end{figure}

\begin{figure}
\centerline{\psfig{figure=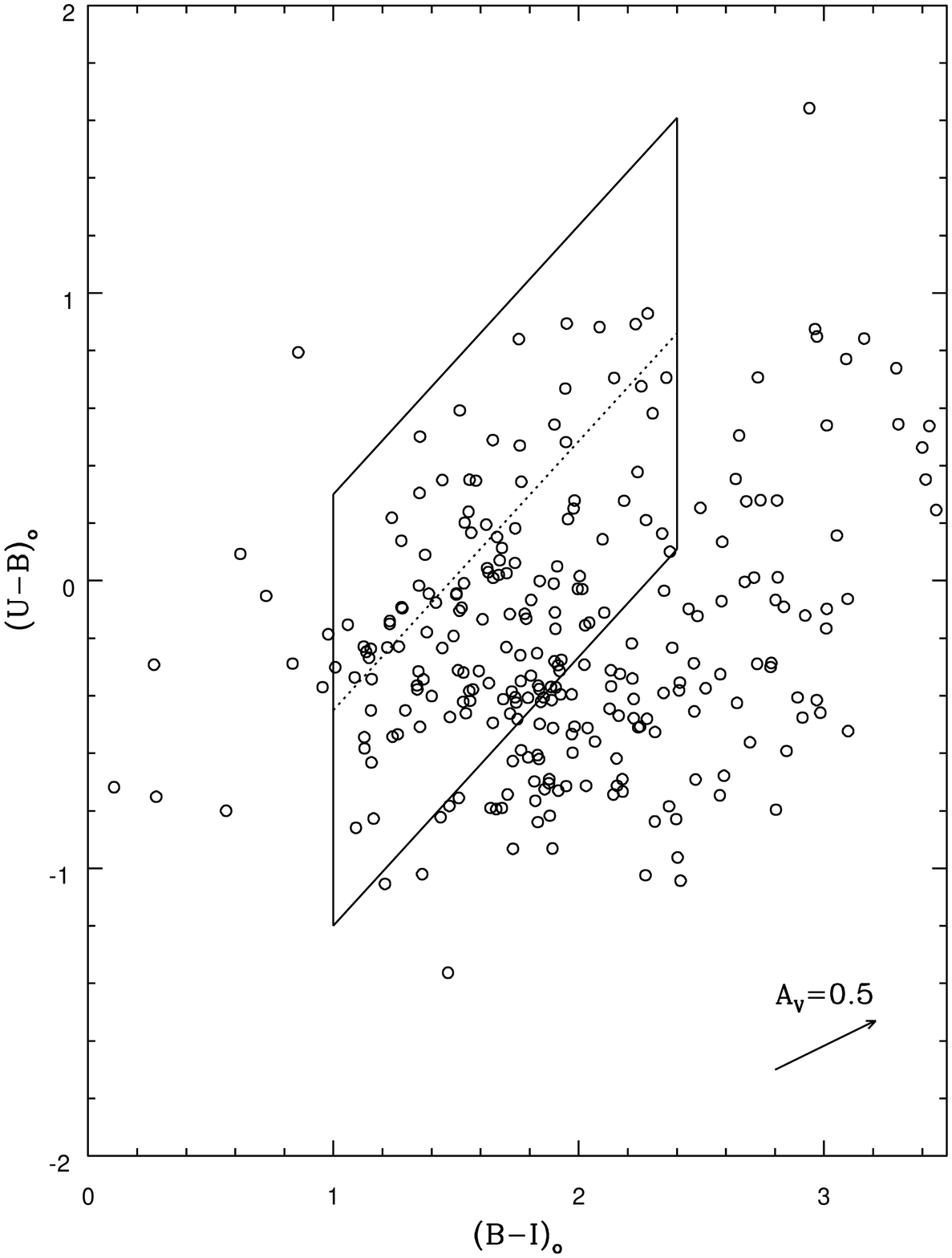,width=0.5\textwidth,angle=0}}
 \caption
{(U--B)$_o$ vs (B--I)$_o$ colour distribution for the candidate
GCs from the large area INT images,
after applying B magnitude cuts. The large number of sources with
(U--B)$_o$ $\sim$ --0.5 are likely to be
contaminating background objects.
}
\label{intubi}
\end{figure}

In order to extend our study beyond the central few
arcminutes of NGC~3379 (i.e. that covered by the Gemini
imaging), we now return to the INT imaging. We restrict candidate
GCs to lie within 5.5$^{'}$ (18 kpc) in galactocentric radius,
as beyond this radius we no longer witness a decline in the
surface density of detected objects (see Fig.~\ref{radial}). 
Thus the majority of objects
interior to this radius appear to be associated with NGC~3379. This
radial selection also ensures there are very few, if any GCs associated
with NGC~3384 (projected separation of 7.2$^{'}$) in our final
object list.

\begin{figure}
\centerline{\psfig{figure=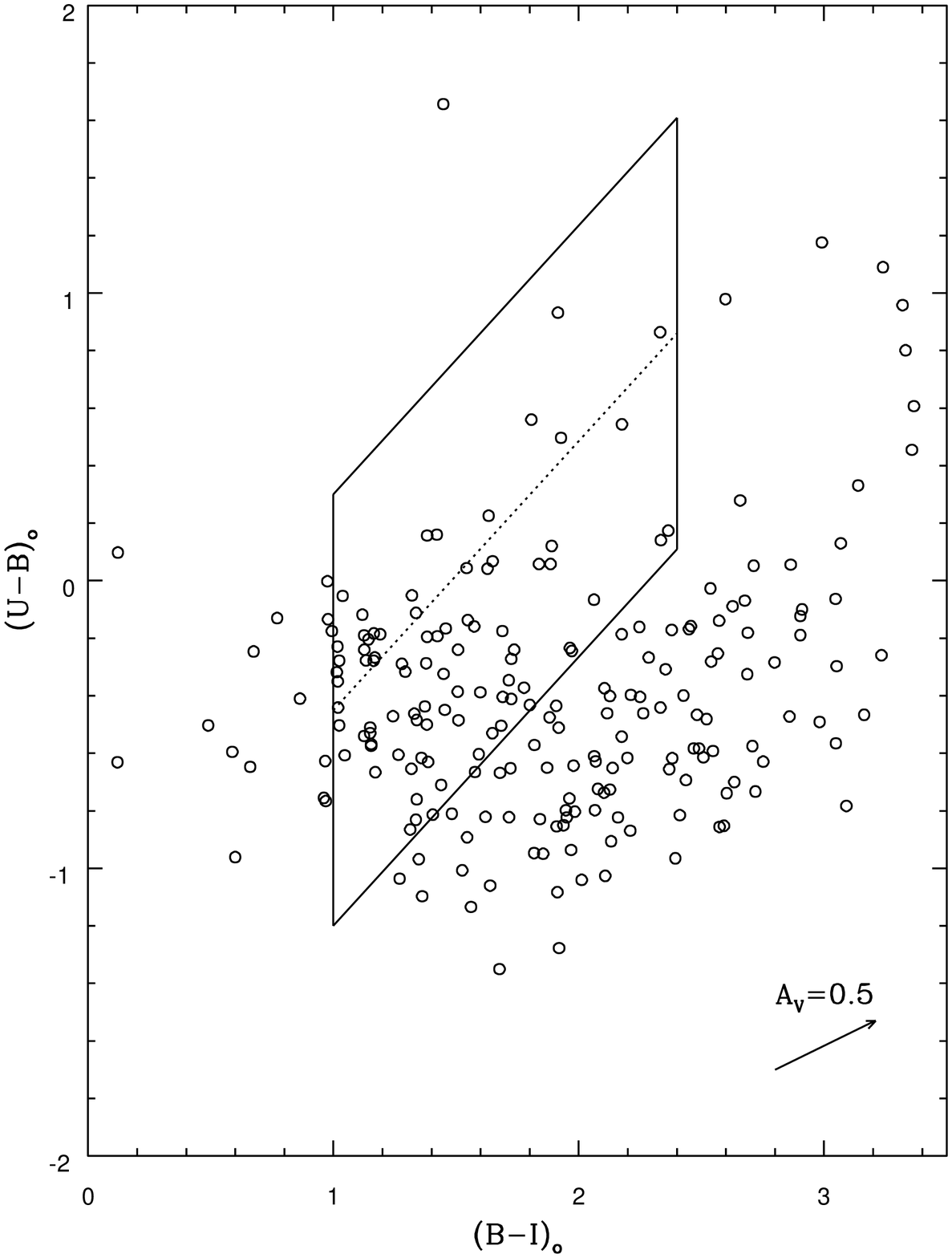,width=0.5\textwidth,angle=0}}
 \caption
{(U--B)$_o$ vs (B--I)$_o$ colour distribution of the background
objects, from an area matched to that for Fig.~\ref{intubi}.
Some of the background objects have colours that fall within the globular
cluster selection region.
}
\label{back}
\end{figure}

We then applied the same B magnitude selection (i.e. 19 $<$ B$_o$
$<$ 23) as
above. The resulting GC candidates in (U--B)$_o$ vs (B--I)$_o$
colour space are shown in Fig.~\ref{intubi}.
Like the Gemini-selected objects, our large area sample reveals a small
number of blue GCs with colours at (U--B)$_o$ $\sim$ 0.1,
(B--I)$_o$ $\sim$ 1.65, but no strong red grouping.

We find 133 candidate GCs within the
colour selected region. The measured magnitudes and positions of
these objects are listed in Appendix A.
How does this final number compare with the
total number of GCs estimated by Harris \& van den Bergh (1981)
of 290 $\pm$ 150 ? Our data have complete radial coverage (with
the possibility of missing a few GCs in the very inner regions) but
clearly under-sample the GC luminosity function. We reach
magnitudes similar to, or slightly brighter than, the expected
turnover magnitude. Thus crudely we expect a factor of 2--2.5$\times$
more GCs than we detected, i.e. 266--333.

Examination of the colour selected
region however suggests that it contains
a number of background objects, i.e. objects
with a wide range of (B--I)$_o$ colours and a mean (U--B)$_o$
colour of about
--0.5. These were not generally seen in the Gemini-selected data,
and are presumably not GCs but background galaxies. So the total number
of GCs in NGC~3379 may be closer to 250 than 300. Both values are
consistent with the low GC specific
frequency found by Harris \& van den Bergh (1981).

\begin{figure}
\centerline{\psfig{figure=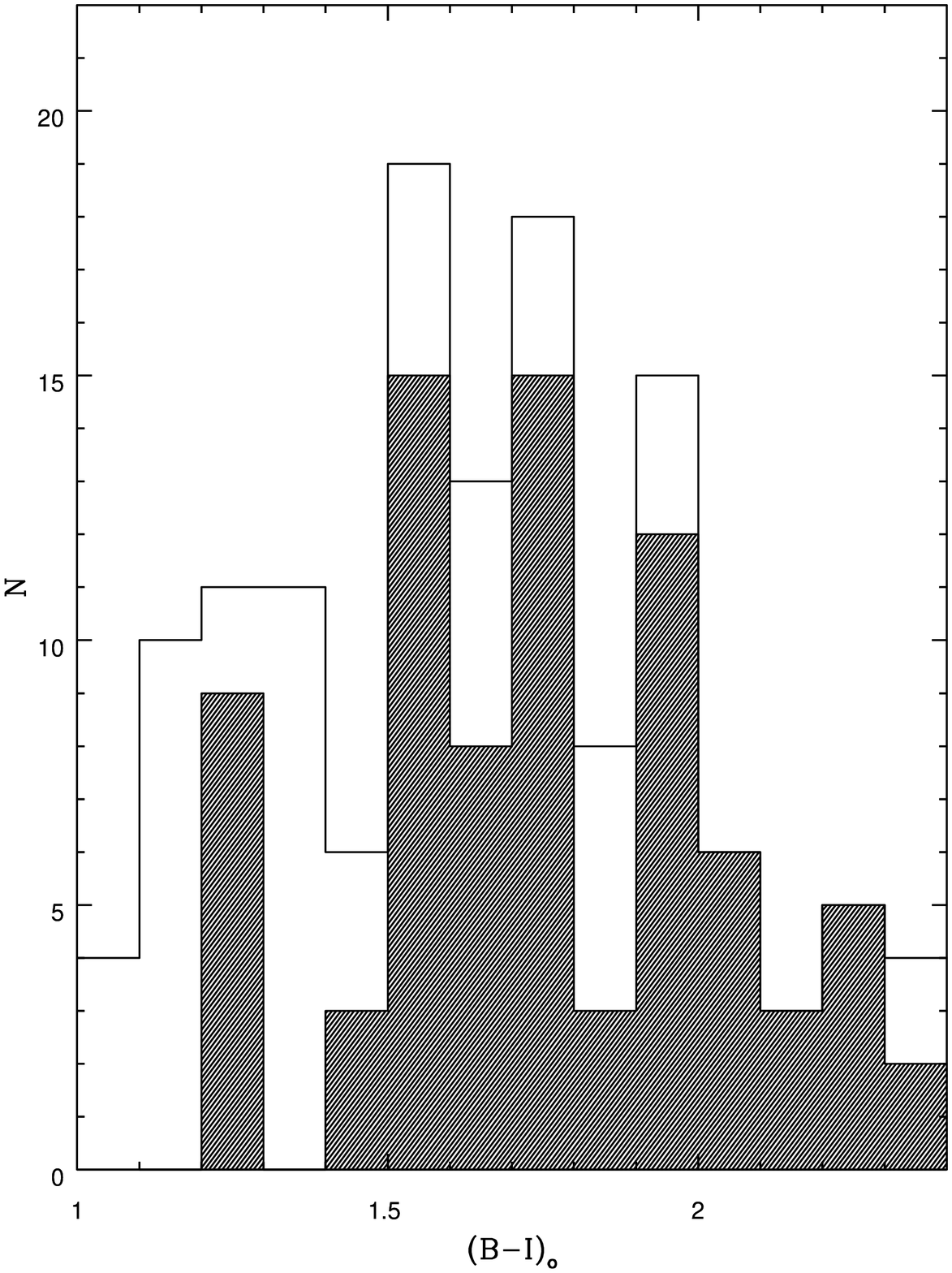,width=0.5\textwidth,angle=0}}
 \caption
{Histogram of (B--I)$_o$ colours for candidate GCs.
The open histogram shows the objects within the colour selection
region. The shaded histogram shows the distribution after
applying a statistical background correction. Both before and after
background correction, the histogram shows evidence for two
subpopulations at (B--I)$_o$ $\sim$ 1.65 and 1.90.
}
\label{inthist}
\end{figure}

In order to further investigate this issue we have defined a
background region of the same central area as studied above but 
located at a galactocentric radius of more than
13.3$^{'}$ (44 kpc). The background
objects, after B magnitude selection, are
shown in Fig.~\ref{back}. Indeed these background objects cover a range in
(B--I)$_o$ with a mean (U--B)$_o$ of about --0.5.
In Fig.~\ref{inthist} we show the (B--I)$_o$ colour distribution for
GC candidates before and after statistical background
subtraction. Two GC subpopulations are revealed with peaks around
(B--I)$_o$ $\sim$ 1.65 and 1.90.
As with the Gemini-selected data, the blue subpopulation is
easily identified in both colour-colour space and in the (B--I)$_o$
distribution. However,
the red subpopulation is only clearly identified in the (B--I)$_o$
distribution.

\section{Globular Cluster Mean Ages and Metallicities}

In order to interpret our data on NGC~3379 we compare it to
similar data for the Milky Way and M31 GC systems. To aid in this
interpretation we have created smoothed
colour  density plots using a Gaussian smoothing kernel. In
Fig.~\ref{dens3} we show the density plots for the MW and M31
(i.e. a smoothed version of Fig.~\ref{double}), with a
background-subtracted
density plot from the large area coverage of NGC~3379 (i.e. Fig. 7
minus Fig. 8).

We first discuss the Milky Way and M31 GC systems.
For the Milky Way, the data reveals the dominant blue
subpopulation at (U--B)$_o$ = 0.04 and (B--I)$_o$ = 1.51, with an
uncertainty in the peak location of $\pm$
0.05.
The red subpopulation is harder to define as it contains
fewer GCs. However a small enhancement can be seen at
(U--B)$_o$ $\sim$ 0.35 and (B--I)$_o$
$\sim$ 1.81.
The distribution for M31 is similar to the Milky Way.
We estimate the dominant blue subpopulation to be at
(U--B)$_o$ = 0.12 and (B--I)$_o$ =
1.62 $\pm$ 0.05.
The red subpopulation appears to have a centre
around (U--B)$_o$ = 0.45 and (B--I)$_o$ = 1.90 $\pm$ 0.05.

An empirical transformation between (U--B)$_o$ and (B--I)$_o$ to
[Fe/H] is given by Barmby \etal (2000). Barmby \etal calculated
these transformations based on spectroscopic metallicities for
$\sim$ 80 Milky Way GCs. Using these transformations we can
calculate `photometric metallicities' based on our colour
estimates. These are summarised in Table 1.
The mean metallicity of the two subpopulations from
spectroscopy are [Fe/H] = --1.59, --0.55 and --1.40, --0.58 for
the Milky Way and M31 respectively (Forbes \etal 2000).
Table 1 shows a good correspondence between our photometric and
the spectroscopic metallicities.

We find that the mean colours (metallicities) of the M31 GC system are
redder (more metal-rich) than that of the Milky Way's GC
system. This is also the situation for the
spectroscopically-defined
metallicities. This is consistent with the mass (luminosity)
of the Milky Way being less than that for M31, as per the galaxy
luminosity - GC metallicity relation (Forbes \& Forte 2001;
Larsen \etal 2001).

\begin{table}
\begin{center}
\renewcommand{\arraystretch}{1.0}
\begin{tabular}{lccc}
\multicolumn{4}{c}{{\bf Table 1.} Globular Cluster Metallicities}\\
\hline
 & Spectra & (U--B)$_o$ & (B--I)$_o$ \\
\hline
Milky Way & & &\\
Metal-poor & --1.59 & --1.75 & --1.73\\
Metal-rich & --0.55 & --0.89 & --0.89\\
\hline
M31 & & &\\
Metal-poor & --1.40 & --1.53 & --1.42\\
Metal-rich & --0.58 & --0.62 & --0.64\\
\hline
NGC~3379 & & &\\
Metal-poor & -- & --1.64 & --1.36\\
Metal-rich & -- & --& --0.6\\
\hline

\end{tabular}
\end{center}
\end{table}

Fig.~\ref{dens3} also includes model tracks from
Brocato \etal (2000) and Maraston (2003, priv. comm.) for a 15
Gyr old single stellar population (SSP). We find that the 15 Gyr old
tracks are closer to the location of the metal-rich subpopulation
than say the 12 Gyr, or younger, model tracks (the metal-poor
subpopulation is equally well fit by a 15 or 12 Gyr old track).
Assuming the difference is purely in the U--B colour, the models
are too blue by 0.1--0.15 magnitudes.
Given a mean age for Milky Way GCs from colour-magnitude diagrams
of around 12 Gyrs (e.g. Salaris \& Weiss 2002), it suggests
that these SSP models are not yet on the correct {\it absolute}
age scale.
The tracks show a range of metallicities for a 15 Gyr old
population. The Milky Way blue
subpopulation has a peak close to a metallicity of [Fe/H] = --1.5, with
the M31 blue GCs between [Fe/H] = --1.5 and --1.0. For both
galaxies, the red subpopulation lies between [Fe/H] = --0.3 and
--1.3.

We have also compared the SSP models of Worthey (1994) with our
data and find that the Worthey models fall well below the
location of the metal-rich subpopulation for any age.
A detailed comparison between the various stellar population models is
beyond the scope of this paper. However, the origin of the
large differences between the U--B colours of the
Worthey models, and those of Brocato \etal
and Maraston, seem to lie in a combination of the theoretical
isochrones adopted (those from Vandenberg 1985; Green \etal 1987
in the case of the Worthey models)
and the conversion between
luminosity/temperature to magnitude/colour in the
observational plane. The Worthey models achieved this
by multiplying observed stellar fluxes by empirical filter
transmission functions, whilst Maraston employ a combination
of empirical and theoretical colour-temperature
relations (see Maraston 1998). As discussed
by Brocato \etal (2000), variations in other model ingredients
such as the IMF slope and low-mass cut-off
do not significantly affect the predicted integrated colours. In any
event, regardless of the specific origin of these discrepancies, we
conclude that the U--B colours of the Worthey models do not accurately
reflect the colours of globular clusters.


Turning now to NGC~3379. The colour density plot for NGC~3379
clearly shows a blue GC subpopulation near (U--B)$_o$ = 0.08 $\pm$
0.05 and (B--I)$_o$ = 1.64 $\pm$ 0.05. The plot also shows some
remaining galaxy contamination with (U--B)$_o$ $\sim$ --0.3. The
colours of these objects are consistent with late-type spirals and
blue compact dwarfs (Schroeder \& Visvanathan 1996).  A red GC
subpopulation is not obvious; it is only revealed when the
distribution in (B--I)$_o$ colour only is examined (see also
Figures 5 and 9). We believe this is due to the larger mean error
in our U--B colours for the metal-rich subpopulation (although the 
presence of an intermediate-aged population could also contribute). 
From the
colour-magnitude diagram (Fig.~\ref{cmd}), it can be seen that the
red subpopulation is fainter on average by about half a magnitude
than the blue one (this is presumably due to additional line
blanketing in the more metal-rich GCs). Larsen \etal (2001) also
found evidence for the red GCs being systematically fainter than
the blue ones.  For their small sample of 21 blue and 24 red GCs,
they estimated V-band turnover magnitudes of 22.57 for the blue
GCs and 23.02 for the red GCs.  For NGC~3379, this results in an
additional colour error of $\sim$ 0.3 mags for the metal-rich
subpopulation. This effect may be spreading out the (fainter)
metal-rich subpopulation. Our B--I colour errors are generally
half those in U--B, and hence are less affected. If we consider
(B--I)$_o$ colour only, i.e. the histograms of Figures 
~\ref{geminihist} (Gemini selected) and ~\ref{inthist} (INT
selected), then we estimate that the red subpopulation has a mean
colour of (B--I)$_o$ $\sim$ 1.9.

The intrinsic U--B and B--I colours of the blue
subpopulation in NGC~3379 are intermediate between those of the
Milky Way and M31 GC systems.
The Barmby \etal (2000) transformation leads to a
photometric metallicity of [Fe/H] = --1.64 $\pm$ 0.14 from (U--B)$_o$ and
--1.36 $\pm$ 0.14 from (B--I)$_o$ for the blue subpopulation.
This is consistent with the metallicity indicated by the Maraston
([Fe/H] $\sim$ --1.35) and Brocato \etal ([Fe/H] $\sim$ --1.5)
15 Gyr SSP tracks.
The photometric metallicity for the red
subpopulation, based on (B--I)$_o$ colour only,
is estimated to be [Fe/H] $\sim$ --0.6. These values
are summarised in Table 1.

The ages of the GC subpopulations in NGC~3379 are less well
constrained by our data. However,
similarities between NGC~3379 and the
Milky Way and M31 GC colours suggests
that the blue subpopulation in all three galaxies has
a similar, old age.
From the colour-magnitude diagram of
Fig.~\ref{cmd},
the red subpopulation in NGC~3379 is fainter in the mean than
the blue GCs. This effectively rules out a young
(i.e. $<$ 3 Gyr) age.

\begin{figure*}
 \caption
{(U--B)$_o$ vs (B--I)$_o$ colour density plot for the Milky Way,
M31 and NGC~3379 globular cluster systems. The solid line shows
the 15 Gyr old model track from Maraston (2003, priv. comm.) with
tick marks for metallicities (from right to left) of [Fe/H] =
0.0, --0.33, --1.35, --2.25. The dashed line shows the 15 Gyr old
model track from Brocato \etal (2000) with
tick marks for metallicities (from right to left) of [Fe/H] =
0.0, --0.5, --1.3, --1.8, --2.3.
The location of the subpopulation mean colours for each galaxy are
shown by a white cross (NGC~3379 does not have a clear red
subpopulation). The open circle indicates the colours of NGC~3379
galaxy at the effective radius. The high density peak at
(U--B)$_o$ $\sim$ --0.3 and (B--I)$_o$ $\sim$ 1.8 is probably due
to faint background galaxies. THIS FIGURE AVAILABLE SEPARATELY.
}
\label{dens3}
\end{figure*}

\section{Conclusions}\label{sec_conc}

Magnitudes and colours, from the Isaac Newton Telescope,
for 133 candidate globular
clusters around NGC~3379 are presented. Our detection rate is
consistent with a low specific frequency.
These candidates, and a
subsample selected from imaging with the Gemini North Telescope,
reveal evidence for a blue and red subpopulation.

We have compared the (U--B)$_o$ vs (B--I)$_o$ colours of the NGC~3379
globular cluster system with that for the Milky Way and M31.
The Milky Way and M31 reveal similar GC colour distributions,
although with slightly different slopes (which may simply be the
result of uncertain reddening corrections). The metal-poor and
metal-rich subpopulations can be seen as a dominant blue and a less
well-defined red peak.
Using the single stellar
population models of Maraston (2003, priv. comm.)
and Brocato \etal (2000), we find that
the mean colours of both subpopulations are best reproduced by
their 15 Gyr
old tracks. Even so, small colour differences between the model
and the measurements exist.
As the mean age of the Milky Way globular clusters
is thought to be closer to
12 Gyrs, it suggests that these model require a relative
age adjustment. The estimated
mean metallicities of the two subpopulations are
very similar to those measured previously from spectroscopy.
We also investigated the models of Worthey (1994), and found they
were unable to
reproduce the (U--B)$_o$ colour of the metal-rich subpopulation
in M31 and the Milky Way for any age.

For NGC~3379 we detect a blue subpopulation with very similar
(U--B)$_o$ and (B--I)$_o$ colours, and presumably age/metallicity,
to that of the Milky Way and M31 globular cluster systems. Thus
the blue GCs in NGC~3379 are consistent with being very old and
with a
mean metallicity of [Fe/H] $\sim$ --1.5.
The red subpopulation is less well-defined, perhaps due to
increased photometric errors, but has a similar mean (B--I)$_o$
colour to the Milky Way and M31 globular cluster systems.
This implies a mean photometric metallicity of [Fe/H] $\sim$ --0.6.

\section{Acknowledgments}\label{sec_ack}

We thank M. Pierce and J. Strader for useful comments. We also
thank A. Terlevich for his help in observing and
the initial data reduction. The data used in this project were
obtained with Isaac Newton Telescope at La Palma Observatory and
the Gemini North Telescope.
This research has made use of the NASA/IPAC Extragalactic
Database (NED) which is operated by the Jet Propulsion
Laboratory, California Institute of Technology, under contract
with the National Aeronautics and Space Administration. \\

\begin{table*}
\begin{center}
\renewcommand{\arraystretch}{1.0}
\begin{tabular}{lccccccc}
\multicolumn{8}{c}{{\bf Table A1.} Candidate Globular Clusters
around NGC~3379}\\
\hline
ID & RA & Dec. & U & B & R & I & B--I\\
   & (J2000) & (J2000) & (mag) & (mag) & (mag) & (mag) & (mag) \\
\hline
001 & 10:48:11.36 & +12:34:32.14 & 20.18 & 20.17 & 18.79 & 18.16 & 2.00\\
002 & 10:48:11.22 & +12:35:42.37 & 21.93 & 22.30 & 20.96 & 20.46 & 1.84\\
003 & 10:48:10.99 & +12:33:50.10 & 20.92 & 21.29 & 20.02 & 19.41 & 1.89\\
004 & 10:48:10.04 & +12:35:09.40 & 21.28 & 21.66 & 20.65 & 19.94 & 1.73\\
005 & 10:48:06.50 & +12:37:57.63 & 22.36 & 21.68 & 20.25 & 19.43 & 2.25\\
006 & 10:48:06.36 & +12:36:10.70 & 22.18 & 22.58 & 21.42 & 20.84 & 1.74\\
007 & 10:48:06.35 & +12:35:33.91 & 22.33 & 22.66 & 21.29 & 20.86 & 1.80\\
008 & 10:48:05.74 & +12:34:17.05 & 21.34 & 21.65 & 20.43 & 19.73 & 1.92\\
009 & 10:48:05.61 & +12:37:15.56 & 23.08 & 22.98 & 21.13 & 20.60 & 2.37\\
010 & 10:48:05.28 & +12:33:07.97 & 21.08 & 21.56 & 20.62 & 20.08 & 1.48\\
011 & 10:48:05.22 & +12:36:48.93 & 21.16 & 21.62 & 20.65 & 20.09 & 1.54\\
012 & 10:48:04.74 & +12:38:16.43 & 22.41 & 23.00 & 21.73 & 21.87 & 1.13\\
013 & 10:48:04.44 & +12:34:12.39 & 21.62 & 21.67 & 20.55 & 20.28 & 1.39\\
014$^{\ast}$ & 10:48:04.33 & +12:35:41.48 & 23.18 & 22.69 & 21.38 & 20.75 & 1.95\\
015$^{\ast}$ & 10:48:04.09 & +12:38:36.88 & 20.54 & 20.85 & 19.85 & 19.34 & 1.51\\
016 & 10:48:03.19 & +12:38:56.00 & 22.83 & 22.69 & 21.24 & 21.41 & 1.28\\
017 & 10:48:03.17 & +12:34:36.48 & 20.32 & 20.66 & 19.92 & 19.57 & 1.09\\
018 & 10:48:02.40 & +12:39:12.83 & 22.53 & 22.95 & 21.73 & 21.39 & 1.56\\
019 & 10:48:01.77 & +12:32:48.29 & 21.77 & 22.02 & 20.80 & 20.19 & 1.83\\
020$^{\ast}$ & 10:48:01.68 & +12:35:59.72 & 21.81 & 21.66 & 20.56 & 19.99 & 1.67\\
021 & 10:48:01.59 & +12:32:39.15 & 22.15 & 22.60 & 21.84 & 21.30 & 1.29\\
022 & 10:48:01.23 & +12:34:20.79 & 23.14 & 22.80 & 21.82 & 21.03 & 1.77\\
023 & 10:48:01.01 & +12:34:16.68 & 23.27 & 22.89 & 21.78 & 20.65 & 2.24\\
024 & 10:48:00.74 & +12:39:26.23 & 21.25 & 21.55 & 20.76 & 20.54 & 1.01\\
025 & 10:48:00.57 & +12:33:41.88 & 22.77 & 23.00 & 21.89 & 21.73 & 1.27\\
026$^{\ast}$ & 10:48:00.28 & +12:32:50.55 & 21.53 & 21.67 & 20.35 & 19.63 & 2.04\\
027$^{\ast}$ & 10:47:58.61 & +12:33:57.32 & 22.85 & 22.50 & 21.72 & 20.94 & 1.55\\
028 & 10:47:58.29 & +12:39:34.80 & 22.29 & 22.38 & 21.63 & 21.10 & 1.28\\
029 & 10:47:57.68 & +12:32:08.49 & 21.71 & 22.25 & 21.48 & 21.01 & 1.24\\
030$^{\ast}$ & 10:47:57.64 & +12:34:29.41 & 23.34 & 22.84 & 21.83 & 21.49 & 1.35\\
031$^{\ast}$ & 10:47:56.56 & +12:33:29.09 & 21.06 & 21.16 & 20.02 & 19.63 & 1.52\\
032 & 10:47:56.51 & +12:32:09.45 & 21.51 & 22.37 & 21.73 & 21.28 & 1.09\\
033 & 10:47:55.19 & +12:30:37.16 & 22.71 & 22.76 & 21.76 & 21.25 & 1.50\\
034 & 10:47:55.16 & +12:30:21.77 & 20.19 & 20.53 & 19.79 & 19.38 & 1.16\\
035 & 10:47:55.13 & +12:32:05.35 & 21.38 & 21.69 & 21.01 & 20.10 & 1.59\\
036$^{\ast}$ & 10:47:54.21 & +12:38:36.91 & 23.08 & 22.78 & 21.98 & 21.43 & 1.35\\
037$^{\ast}$ & 10:47:53.97 & +12:36:32.31 & 22.36 & 21.69 & 20.47 & 19.75 & 1.95\\
038$^{\ast}$ & 10:47:53.91 & +12:34:54.28 & 23.25 & 22.66 & 21.50 & 21.14 & 1.52\\
039 & 10:47:53.81 & +12:31:09.03 & 20.05 & 20.28 & 19.49 & 19.16 & 1.12\\
040 & 10:47:53.75 & +12:32:56.42 & 20.60 & 20.89 & 19.66 & 18.98 & 1.91\\
041$^{\ast}$ & 10:47:53.74 & +12:33:48.33 & 22.47 & 22.27 & 21.09 & 20.65 & 1.62\\
042$^{\ast}$ & 10:47:53.48 & +12:34:55.67 & 21.88 & 21.67 & 20.68 & 20.14 & 1.53\\
043$^{\ast}$ & 10:47:53.32 & +12:34:14.15 & 22.61 & 22.62 & 21.44 & 20.72 & 1.90\\
044$^{\ast}$ & 10:47:53.29 & +12:35:05.76 & 23.23 & 22.53 & 21.01 & 20.38 & 2.14\\
045$^{\ast}$ & 10:47:52.46 & +12:36:00.51 & 23.14 & 22.25 & 20.75 & 20.02 & 2.23\\
046 & 10:47:51.71 & +12:32:52.10 & 22.78 & 22.89 & 21.58 & 20.79 & 2.10\\
047 & 10:47:51.37 & +12:40:08.89 & 22.24 & 22.73 & 21.56 & 21.08 & 1.65\\
048$^{\ast}$ & 10:47:50.99 & +12:35:49.78 & 20.78 & 20.75 & 19.53 & 19.05 & 1.71\\
049 & 10:47:50.75 & +12:30:22.69 & 21.93 & 22.17 & 21.24 & 20.73 & 1.44\\
050$^{\ast}$ & 10:47:50.68 & +12:35:30.07 & 21.49 & 21.31 & 20.13 & 19.57 & 1.74\\
051$^{\ast}$ & 10:47:50.56 & +12:35:32.43 & 21.57 & 21.35 & 20.15 & 19.39 & 1.96\\
052$^{\ast}$ & 10:47:50.47 & +12:35:19.56 & 22.49 & 22.44 & 21.37 & 20.53 & 1.91\\
053$^{\ast}$ & 10:47:50.43 & +12:34:37.76 & 22.18 & 21.83 & 20.65 & 20.39 & 1.44\\
054$^{\ast}$ & 10:47:50.42 & +12:33:49.17 & 22.53 & 22.04 & 20.84 & 20.39 & 1.65\\
055 & 10:47:50.26 & +12:32:22.91 & 21.59 & 22.41 & 21.30 & 21.25 & 1.16\\
056 & 10:47:50.10 & +12:31:06.72 & 22.42 & 22.76 & 21.75 & 21.39 & 1.37\\
057$^{\ast}$ & 10:47:49.67 & +12:34:33.57 & 20.22 & 20.11 & 18.88 & 18.42 & 1.69\\
058$^{\ast}$ & 10:47:49.66 & +12:34:11.20 & 22.82 & 22.24 & 20.80 & 19.94 & 2.30\\
059$^{\ast}$ & 10:47:48.96 & +12:35:23.11 & 21.50 & 21.66 & 20.68 & 20.43 & 1.23\\
\hline

\end{tabular}
\end{center}
\end{table*}

\begin{table*}
\begin{center}
\renewcommand{\arraystretch}{1.0}
\begin{tabular}{lccccccc}
\multicolumn{8}{c}{{\bf Table A1.} Candidate Globular Clusters
around NGC~3379}\\
\hline
ID & RA & Dec. & U & B & R & I & B--I\\
   & (J2000) & (J2000) & (mag) & (mag) & (mag) & (mag) & (mag) \\
\hline
060 & 10:47:48.60 & +12:37:03.83 & 21.72 & 21.77 & 20.79 & 20.27 & 1.50\\
061 & 10:47:48.59 & +12:35:32.40 & 20.98 & 21.15 & 20.25 & 19.77 & 1.38\\
062$^{\ast}$ & 10:47:48.20 & +12:35:45.07 & 21.59 & 21.70 & 20.45 & 19.98 & 1.72\\
063$^{\ast}$ & 10:47:48.00 & +12:35:07.78 & 21.29 & 21.67 & 20.38 & 20.10 & 1.57\\
064$^{\ast}$ & 10:47:47.91 & +12:34:38.75 & 21.94 & 22.34 & 20.86 & 20.94 & 1.40\\
065$^{\ast}$ & 10:47:47.90 & +12:36:22.85 & 23.01 & 22.79 & 21.82 & 21.55 & 1.24\\
066$^{\ast}$ & 10:47:47.87 & +12:35:05.17 & 21.61 & 21.54 & 20.41 & 19.86 & 1.68\\
067$^{\ast}$ & 10:47:47.80 & +12:34:15.41 & 22.84 & 21.94 & 20.51 & 19.99 & 1.95\\
068$^{\ast}$ & 10:47:47.78 & +12:36:41.23 & 21.62 & 21.90 & 20.73 & 20.00 & 1.90\\
069 & 10:47:47.40 & +12:37:19.00 & 22.38 & 22.63 & 21.81 & 20.87 & 1.76\\
070$^{\ast}$ & 10:47:47.30 & +12:34:05.47 & 22.86 & 22.59 & 21.39 & 20.60 & 1.98\\
071 & 10:47:46.85 & +12:36:52.44 & 21.85 & 22.21 & 21.50 & 20.87 & 1.34\\
072 & 10:47:46.78 & +12:31:37.35 & 21.92 & 22.34 & 22.11 & 20.81 & 1.53\\
073 & 10:47:46.48 & +12:32:34.57 & 22.12 & 22.50 & 21.38 & 21.16 & 1.34\\
074 & 10:47:46.30 & +12:31:19.99 & 21.95 & 21.86 & 20.73 & 20.49 & 1.37\\
075 & 10:47:46.22 & +12:38:49.41 & 20.19 & 20.61 & 19.33 & 18.87 & 1.75 \\
076$^{\ast}$ & 10:47:45.79 & +12:34:48.40 & 21.99 & 21.99 & 20.63 & 20.15 & 1.84\\
077$^{\ast}$ & 10:47:45.65 & +12:34:51.91 & 21.73 & 22.00 & 20.79 & 20.07 & 1.93\\
078 & 10:47:45.54 & +12:40:08.99 & 20.81 & 21.32 & 20.42 & 19.97 & 1.35\\
079 & 10:47:45.44 & +12:31:33.17 & 21.57 & 21.80 & 21.03 & 20.58 & 1.22\\
080 & 10:47:45.19 & +12:39:57.83 & 21.40 & 21.65 & 20.89 & 20.52 & 1.13\\
081 & 10:47:43.76 & +12:30:45.96 & 20.07 & 20.18 & 19.09 & 18.40 & 1.78\\
082$^{\ast}$ & 10:47:43.67 & +12:33:34.16 & 22.32 & 22.47 & 21.26 & 21.41 & 1.06\\
083$^{\ast}$ & 10:47:43.55 & +12:37:04.17 & 22.20 & 22.34 & 21.35 & 21.11 & 1.23\\
084$^{\ast}$ & 10:47:43.34 & +12:36:46.37 & 22.58 & 22.34 & 21.09 & 20.79 & 1.55\\
085$^{\ast}$ & 10:47:42.25 & +12:34:47.47 & 22.51 & 22.49 & 21.30 & 20.82 & 1.67\\
086 & 10:47:42.14 & +12:35:08.62 & 21.99 & 22.62 & 21.87 & 21.46 & 1.15\\
087$^{\ast}$ & 10:47:42.11 & +12:36:10.11 & 22.31 & 22.34 & 21.14 & 20.34 & 2.00\\
088$^{\ast}$ & 10:47:41.84 & +12:35:40.68 & 23.10 & 22.26 & 21.12 & 20.50 & 1.76\\
089$^{\ast}$ & 10:47:41.30 & +12:36:50.81 & 22.91 & 22.87 & 21.80 & 21.24 & 1.63\\
090 & 10:47:40.91 & +12:31:37.45 & 21.54 & 22.08 & 21.51 & 20.81 & 1.26\\
091$^{\ast}$ & 10:47:40.82 & +12:34:05.12 & 22.01 & 22.47 & 21.29 & 20.75 & 1.72\\
092 & 10:47:40.80 & +12:39:42.82 & 21.54 & 21.86 & 20.68 & 20.33 & 1.53\\
093 & 10:47:40.73 & +12:32:17.44 & 19.61 & 19.70 & 18.79 & 18.42 & 1.28\\
094 & 10:47:40.61 & +12:38:57.95 & 19.74 & 19.93 & 18.98 & 18.44 & 1.49\\
095$^{\ast}$ & 10:47:40.61 & +12:35:22.13 & 20.50 & 20.60 & 19.59 & 19.09 & 1.51\\
096$^{\ast}$ & 10:47:40.60 & +12:35:53.78 & 20.94 & 20.91 & 19.87 & 19.28 & 1.63\\
097 & 10:47:40.51 & +12:37:22.36 & 22.43 & 22.26 & 21.05 & 20.70 & 1.56\\
098 & 10:47:40.03 & +12:34:36.04 & 22.68 & 22.83 & 21.57 & 20.80 & 2.03\\
099$^{\ast}$ & 10:47:39.71 & +12:33:58.78 & 22.64 & 22.17 & 20.98 & 20.41 & 1.76\\
100$^{\ast}$ & 10:47:39.70 & +12:32:26.12 & 20.82 & 20.76 & 19.55 & 19.02 & 1.74\\
101 & 10:47:39.47 & +12:35:37.17 & 21.30 & 21.75 & 20.85 & 20.60 & 1.15\\
102 & 10:47:39.44 & +12:39:37.58 & 22.54 & 22.26 & 20.96 & 20.08 & 2.18\\
103 & 10:47:39.33 & +12:30:53.43 & 22.11 & 21.90 & 20.33 & 19.63 & 2.27\\
104 & 10:47:39.32 & +12:39:42.28 & 21.91 & 22.14 & 21.15 & 20.43 & 1.71\\
105 & 10:47:39.24 & +12:35:35.50 & 21.49 & 21.91 & 20.87 & 20.21 & 1.69\\
106$^{\ast}$ & 10:47:39.24 & +12:35:01.26 & 23.30 & 22.95 & 21.53 & 21.37 & 1.58\\
107 & 10:47:39.01 & +12:36:45.95 & 22.66 & 22.52 & 21.37 & 20.42 & 2.10\\
108 & 10:47:38.98 & +12:35:26.48 & 22.84 & 22.68 & 21.10 & 20.34 & 2.34\\
109 & 10:47:38.89 & +12:39:38.32 & 22.50 & 22.50 & 21.91 & 20.97 & 1.53\\
110 & 10:47:38.60 & +12:35:57.03 & 22.01 & 22.42 & 21.39 & 20.62 & 1.79\\
111 & 10:47:38.31 & +12:31:53.19 & 21.45 & 21.44 & 20.28 & 19.79 & 1.65\\
112$^{\ast}$ & 10:47:38.28 & +12:35:10.56 & 21.96 & 22.44 & 21.17 & 20.69 & 1.75\\
113 & 10:47:38.21 & +12:39:28.19 & 20.90 & 20.98 & 20.00 & 19.56 & 1.42\\
114 & 10:47:37.84 & +12:31:04.88 & 22.59 & 22.70 & 21.19 & 20.80 & 1.90\\
115 & 10:47:37.76 & +12:33:33.40 & 21.32 & 21.07 & 19.75 & 19.09 & 1.98\\
116$^{\ast}$ & 10:47:37.66 & +12:34:14.78 & 21.80 & 21.82 & 21.32 & 20.47 & 1.35\\
117 & 10:47:37.49 & +12:38:56.74 & 22.23 & 22.77 & 22.28 & 21.65 & 1.13\\
118 & 10:47:37.34 & +12:36:02.54 & 21.61 & 21.96 & 20.66 & 20.33 & 1.63\\
119$^{\ast}$ & 10:47:37.14 & +12:35:21.82 & 22.00 & 22.03 & 20.80 & 20.02 & 2.01\\
120 & 10:47:36.91 & +12:37:21.02 & 21.81 & 20.93 & 19.46 & 18.84 & 2.09\\
\hline

\end{tabular}
\end{center}
\end{table*}

\begin{table*}
\begin{center}
\renewcommand{\arraystretch}{1.0}
\begin{tabular}{lccccccc}
\multicolumn{8}{c}{{\bf Table A1.} Candidate Globular Clusters
around NGC~3379}\\
\hline
ID & RA & Dec. & U & B & R & I & B--I\\
   & (J2000) & (J2000) & (mag) & (mag) & (mag) & (mag) & (mag) \\
\hline
121 & 10:47:36.35 & +12:34:41.33 & 20.08 & 20.35 & 19.57 & 19.20 & 1.14\\
122 & 10:47:36.06 & +12:35:53.02 & 22.10 & 22.23 & 21.08 & 20.63 & 1.61\\
123 & 10:47:36.05 & +12:39:12.20 & 21.92 & 21.22 & 19.53 & 18.86 & 2.36\\
124 & 10:47:35.89 & +12:30:45.05 & 21.96 & 22.03 & 20.77 & 20.22 & 1.81\\
125 & 10:47:35.35 & +12:38:47.74 & 21.49 & 21.81 & 21.25 & 20.46 & 1.35\\
126$^{\ast}$ & 10:47:34.01 & +12:33:27.75 & 21.21 & 20.29 & 18.66 & 18.00 & 2.28\\
127$^{\ast}$ & 10:47:33.83 & +12:34:43.89 & 23.07 & 22.53 & 21.17 & 20.62 & 1.90\\
128 & 10:47:33.62 & +12:34:38.74 & 21.16 & 21.33 & 20.14 & 19.42 & 1.91\\
129 & 10:47:33.51 & +12:33:08.65 & 21.03 & 21.38 & 20.18 & 19.62 & 1.76\\
130 & 10:47:32.63 & +12:36:26.46 & 21.43 & 21.67 & 21.13 & 20.51 & 1.15\\
131 & 10:47:32.27 & +12:36:59.72 & 20.95 & 21.09 & 19.91 & 19.30 & 1.79\\
132 & 10:47:31.83 & +12:36:56.07 & 21.87 & 22.25 & 21.08 & 20.70 & 1.55\\
133 & 10:47:31.44 & +12:35:53.33 & 20.07 & 20.45 & 19.29 & 18.61 & 1.84\\
\hline

\end{tabular}
\end{center}
Notes:$^{\ast}$ = Globular cluster present in the
Gemini selected object lists. Galaxy centre is RA = 10:47:49.6, Dec. =
+12:34:55 (J2000).
\end{table*}

\end{document}